\documentclass[12pt,a4paper]{iopart}

\pdfoutput=1

\usepackage{graphicx}
\usepackage{latexsym}
\usepackage{iopams}
\usepackage{bbm}
\usepackage{color}

\renewcommand{\P}{{\mathbb{P}}}

\newcommand{\R}{{\mathbb{R}}}
\newcommand{\E}{{\mathbb{E}}}
\newcommand{\1}{{\mathbbm{1}}}
\newcommand{\feta}{{\boldsymbol{\eta}}}
\newcommand{\Jcal}{{\mathcal{J}}}

\begin{document}

\title{Dynamics of non-Markovian exclusion processes}

\author{Diana Khoromskaia$^1$, Rosemary J. Harris$^2$, Stefan Grosskinsky$^{1,3}$}

\address{$^1$ Centre for Complexity Science, University of Warwick, Coventry CV4 7AL, UK}
\address{$^2$ School of Mathematical Sciences, Queen Mary University of London, London E1~4NS, UK}
\address{$^3$ Mathematics Institute, University of Warwick, Coventry CV4 7AL, UK}
\ead{\mailto{D.Khoromskaia@warwick.ac.uk}, \mailto{rosemary.harris@qmul.ac.uk}, \mailto{s.w.grosskinsky@warwick.ac.uk}}

\begin{abstract}
Driven diffusive systems are often used as simple discrete models of collective transport phenomena in physics, biology or social sciences. Restricting attention to one-dimensional geometries, the asymmetric simple exclusion process (ASEP) plays a paradigmatic role to describe noise-activated driven motion of entities subject to an excluded volume interaction and many variants have been studied in recent years. While in the standard ASEP the noise is Poissonian and the process is therefore Markovian, in many applications the statistics of the activating noise has non-standard distribution with possible memory effects resulting from internal degrees of freedom or external sources. This leads to temporal correlations and can significantly affect the shape of the current-density relation as has been studied recently for a number of scenarios.  In this paper we report a general framework to derive the fundamental diagram of ASEPs driven by non-Poissonian noise by using effectively only two simple quantities, viz., the mean residual lifetime of the jump distribution and a suitably defined temporal correlation length.  We corroborate our results by detailed numerical studies for various noise statistics under periodic boundary conditions and discuss how our approach can be applied to more general driven diffusive systems.
\end{abstract}

\maketitle


\section{Introduction}

The totally asymmetric simple exclusion process (TASEP) is one of the most studied models in non-equilibrium statistical mechanics and a paradigm for driven diffusive collective transport. 
In this stochastic process particles hop uni-directionally on a one-dimensional lattice subject to the constraint that each site can contain only a single particle.
Originally introduced to model kinetics of biopolymerization \cite{MacDonald1968}, variants have since been used to describe many different biological, physical and socioeconomic transport processes~\cite{Chowdhury2005,Neri2013, Stinchcombe13, Schadschneider08} and a rich theoretical literature has also developed (see e.g.,~\cite{Spitzer1970,Derrida1998}).  

In the standard continuous-time TASEP the dynamics are Markovian and the waiting times between jump attempts of a particle have an exponential distribution. The times of attempted jumps then form a Poisson process and one can think of Poisson clocks that ring after exponential times to indicate the next jump attempt. Depending on the application, these clocks can either be attached to particles (such as in the case of molecular motors) or attached to lattice sites or bonds (such as for networks of server queues). As long as the processes all have the same rate, these two interpretations are equivalent in the Markovian case.  

\begin{figure}%
\mbox{\includegraphics[height=0.28\textwidth]{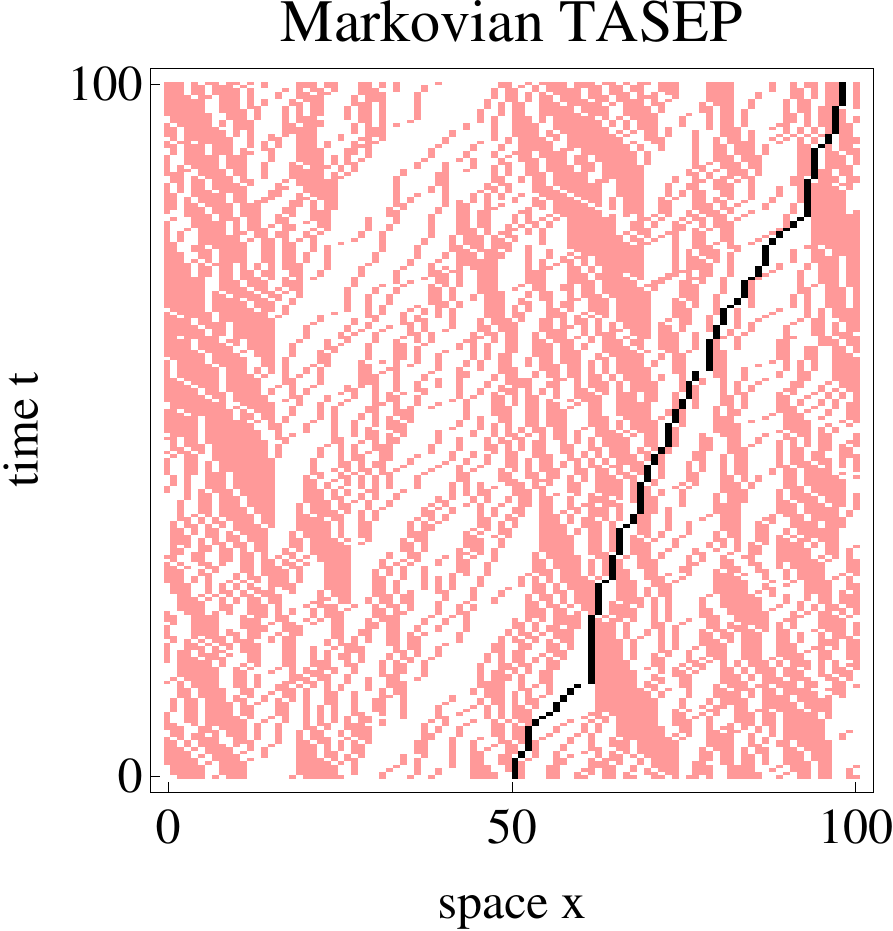}\ \includegraphics[height=0.28\textwidth]{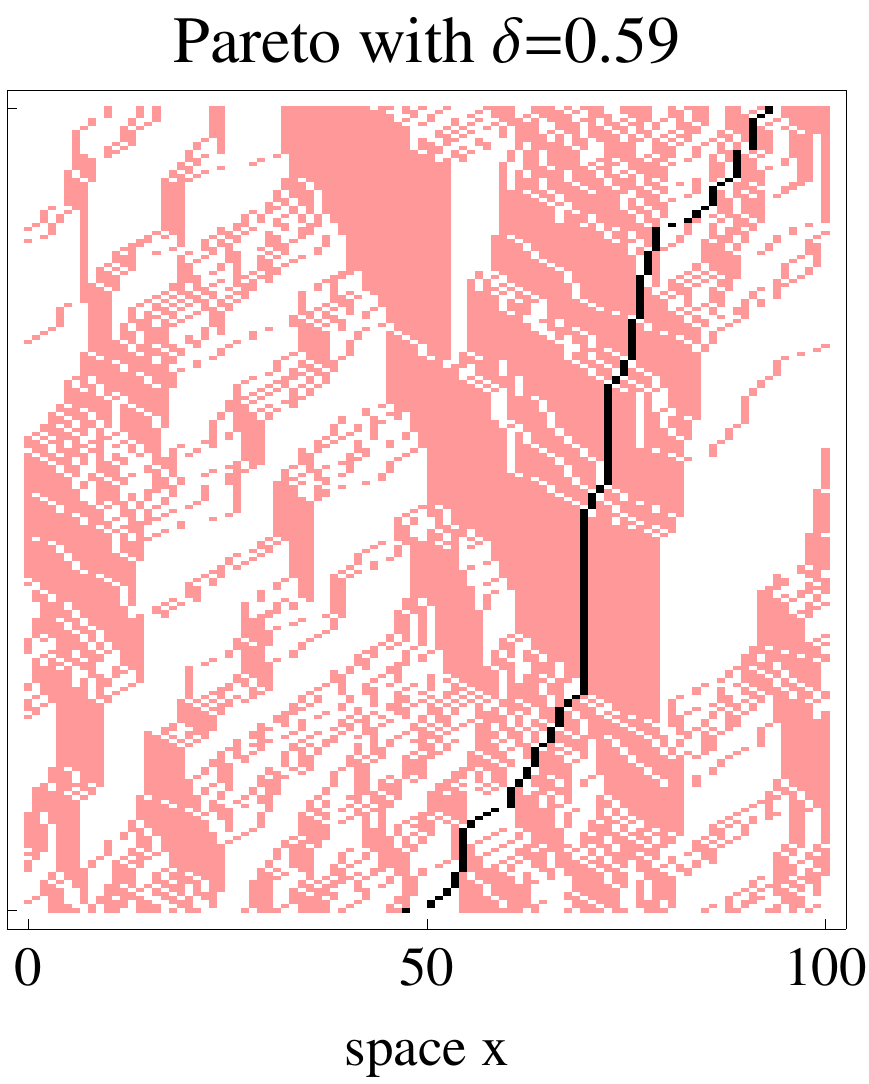}\ \includegraphics[height=0.28\textwidth]{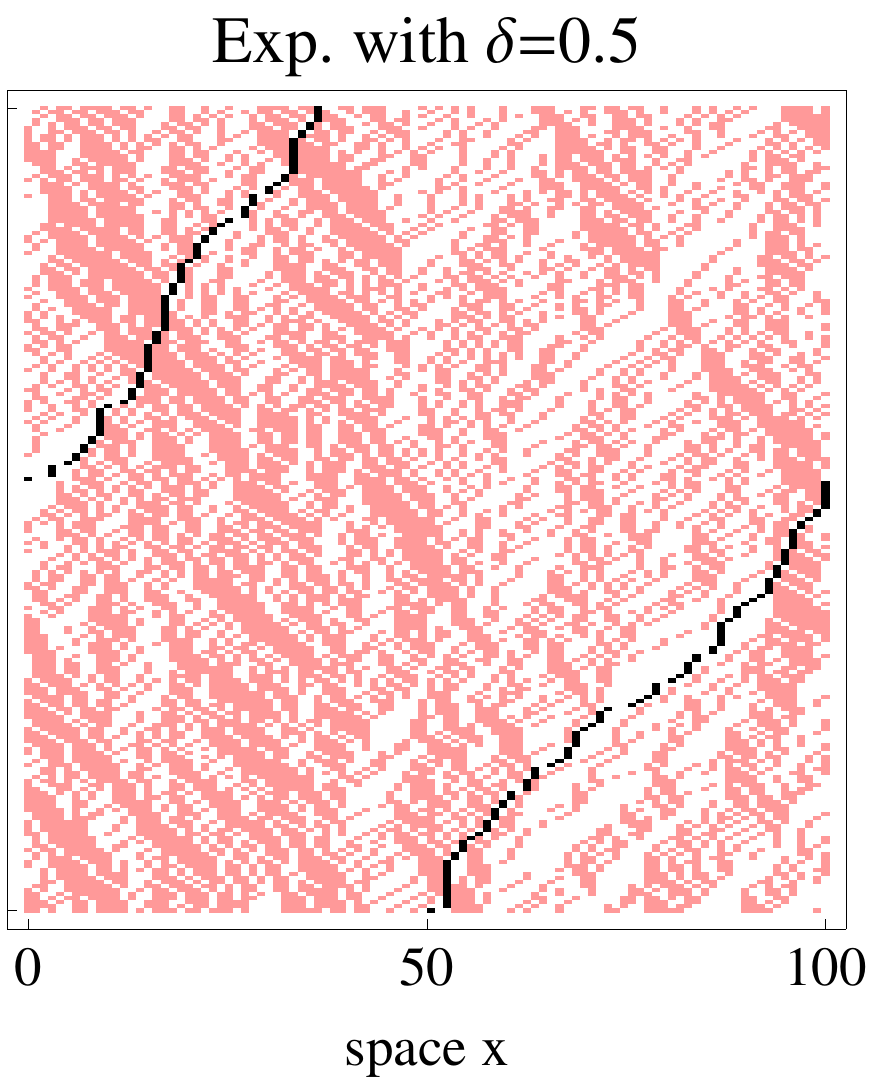}\ \includegraphics[height=0.28\textwidth]{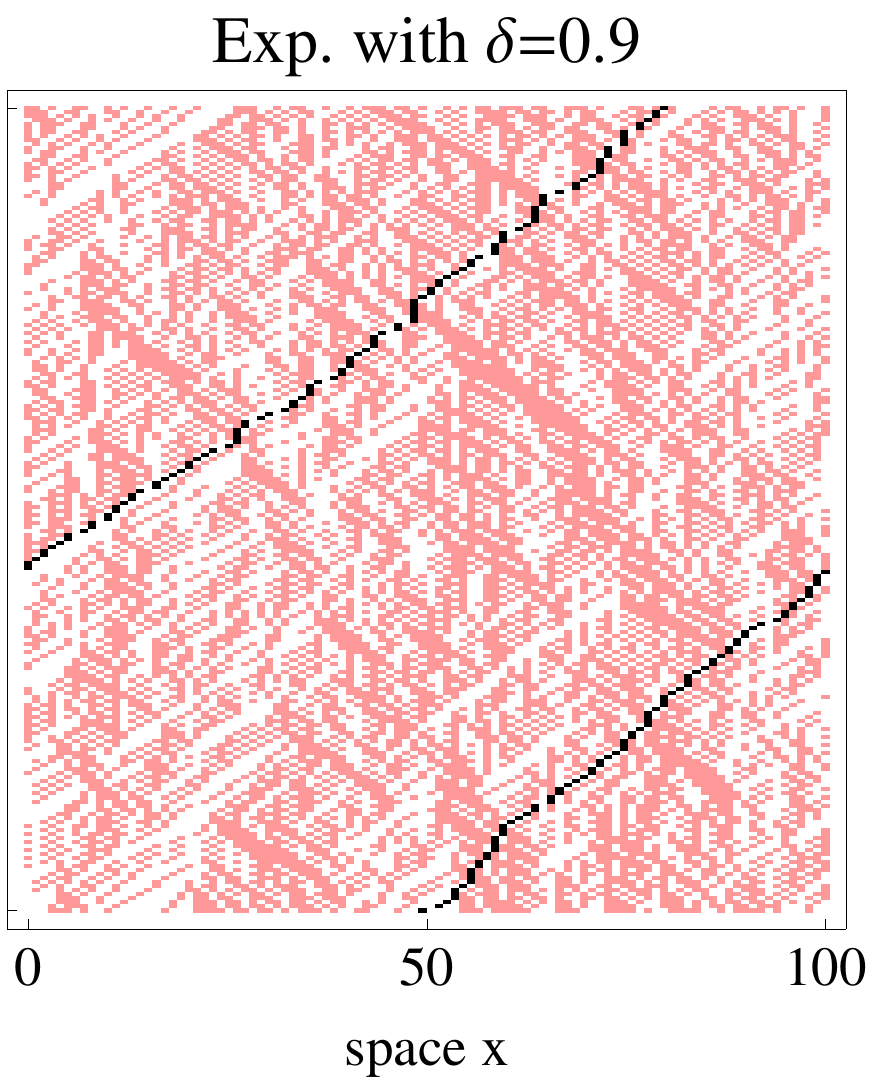}}
\caption{Time evolution of the TASEP at density $\rho =1/2$ with a tagged particle shown in black. The Pareto law (cf.~(\ref{eq:fgamma})) is chosen such that a single, free particle would have the same average speed as in the Markovian case (cf.~(\ref{eq:result1})), while enhanced blockages due to the heavy tail distribution can slow down the tagged particle. More regular waiting time distributions with delayed exponential (cf.~(\ref{eq:fdelta})) lead to correlated motion of the particles and an enhanced current.
}%
\label{fig:tagged}
\end{figure}

We consider, in this paper, the effect of generalized waiting-time distributions with finite mean (fixed to unity throughout). To avoid degeneracies, we assume that the waiting times have non-zero variance 
and that the associated renewal process of attempted jump times is ergodic. An example of such dynamics has been studied recently in \cite{Concannon2011} where non-Poisson clocks with heavy-tailed waiting time distributions of Pareto type were attached to particles. For heavy tails with infinite variance, very long waiting times can be sampled leading to a condensation phenomenon where particles form a macroscopic jam.  In particular, it was observed in~\cite{Concannon2011} that an increase in the coefficient of variation of the waiting time distribution leads to a reduced stationary current.  In contrast, in the present contribution we are chiefly interested in cases where the variation is smaller than for the exponential distribution, leading to correlated stationary states and increased currents as illustrated in figure~\ref{fig:tagged}. This is important to understand collective transport dynamics of motors or particles with internal degrees of freedom leading to non-exponential waiting times, see e.g.,~\cite{cross04,nishinari05,Pinkoviezky2013,Ciandrini2013} and note that the last of these works provides numerical evidence for an increase in maximal current.

Specifically we focus on studying the fundamental diagrams of such non-Markovian models, i.e., the stationary current as a function of the particle density for periodic boundary conditions in the limit of large system size. This is the most important characteristic of a driven diffusive system which essentially fully determines its behaviour on macroscopic scales through conservation laws (see, e.g.,~\cite{Bertini14}).  Furthermore, the well-known phase diagrams for open boundary systems are determined by steady state selection from the fundamental diagram \cite{Popkov00}. An exact calculation requires knowledge of the stationary state, which is not available for general non-Markovian versions of TASEP. However, our main result is that, to a very good approximation, the fundamental diagram is determined by three characteristics of the waiting time distribution: the mean (set to unity), the mean of the residual waiting time distribution (which fixes the slope of the fundamental diagram at high and low densities and is essentially equivalent to the variance of the distribution) and a third order moment of the distribution (which characterizes the deviation from the maximal possible current in the system).

We have tested this claim for a large class of different waiting time distributions, including $\Gamma$-distributions and Pareto distributions, as well as uniform and simple bimodal distributions for which we do not present data here. Our main example for the purpose of a simple presentation is a family of exponential distributions with a delay time $\delta\in [0,1]$, which has a characteristic unimodal structure and interpolates between the Markovian case ($\delta =0$) and the deterministic case ($\delta =1$) where all hop attempts are fully synchronized. Our approach applies to non-Poissonian jump processes being attached to particles, as well as to lattice sites. The latter exhibits a particle-hole symmetry and is therefore simpler to study, so we illustrate our approach for this case (i.e., clocks fixed to sites) and then explain how it can be modified to study the particle-based model.

In a broader context, there are also discrete-time versions of the TASEP which are often used in models of pedestrian or vehicular traffic and are known to lead to similar correlations and modified fundamental diagrams. 
In these models all particles try to update their position after each time step in an order determined by the current update scheme, e.g., parallel update or shuffled update \cite{Rajewsky1998}.  Different implementations lead to different dynamics but often the update rules are artificial and are chosen essentially ad hoc in order to introduce observed correlations and produce realistic fundamental diagrams (see e.g., \cite{schadschneider}).  Our approach provides a natural framework to achieve similar fundamental diagrams in the often more realistic case of continuous-time dynamics.   In particular, it turns out that one limiting case of the particle-based model is the frozen shuffle update \cite{Appert-Rolland2011} where the update order is set randomly in the beginning of the simulation and remains fixed.   This furnishes a theoretical explanation of the approach of the fundamental diagram to an asymmetric triangular shape. 

We conclude this introductory section by remarking that although single-particle non-Markovian dynamics (as relevant for describing the dynamics of individual motors~\cite{Santos2005,Linden2007}) has been studied within the mathematical framework of renewal processes or continuous-time random walks~\cite{Montroll84,Qian06}, there is to our knowledge no general analytical description for non-Markovian many-particle systems and we thus hope our results will be of wider interest. One similar non-Markovian model worth mentioning here is the recently-studied zero-range process with an additional stochastic activity variable attached to each site \cite{hirschberg09,Hirschberg2012} -- this can also be mapped to a type of non-Markovian TASEP. Non-Markovian dynamics are also of recent interest in dynamics of interacting neurons \cite{galves13} and in the context of polymer growth \cite{sokolovski}. The rest of the paper is structured as follows.  After introducing the framework in section~\ref{s:def}, we present our main results in section~\ref{s:results} before discussing generalizations and further questions in section~\ref{s:discussion}.

\section{Definitions and notation\label{s:def}}

We consider a one-dimensional lattice $\Lambda$ of $L$ sites with periodic boundary conditions, and denote a configuration of the TASEP by
\begin{equation}
\feta =(\eta_x :x\in\Lambda )\ ,\quad \mbox{where}\quad\eta_x \in\{ 0,1\}
\label{eq:eta}
\end{equation}
corresponds to the absence or presence of a particle, respectively. In the site-based model, we associate independent jump processes to sites $x$, each represented by an increasing sequence of positive times
\begin{equation}
\tau_x =(\tau_x^1 ,\tau_x^2 ,\ldots )\ ,\quad\mbox{where}\quad \tau_x^k >0
\label{eq:renewal}
\end{equation}
denotes the time of the $k$-th jump attempt at site $x$. We assume the $\tau_x$ to be renewal processes (see e.g., \cite{KarlinTaylor}), so that the waiting times
\begin{equation}
T_x^k :=\tau_x^k -\tau_x^{k-1} >0
\label{eq:dwell}
\end{equation}
are independent and identically distributed random variables; we denote a generic copy by $T$ and its distribution by
\begin{equation}
F(t):=\P (T \leq t)\quad\mbox{with associated PDF}\quad f(t)=F'(t)\ .
\label{eq:dwelldist}
\end{equation}
Our main example throughout is a delayed exponential distribution
\begin{equation}
f_\delta (t)=\frac{1}{1-\delta}\, e^{-(t-\delta)/(1-\delta )}\1_{[\delta ,\infty )} (t)\ ,
\label{eq:fdelta}
\end{equation}
with delay $\delta\in [0,1)$. Note that for $\delta =0$ waiting times are simply exponentially distributed (Markovian case), and that we choose the exponential tail such that the mean $\langle T\rangle =1$ for all $\delta\in [0,1)$. For $\delta\to 1$, the PDF in (\ref{eq:fdelta}) converges to $f_1 (t)=\delta (t-1)$ which corresponds to a deterministic limit $T=1$. 
So this family interpolates between the Markovian and the deterministic case. As further examples we also consider Gamma and Pareto distributed waiting times
\begin{eqnarray}
f_a^\Gamma (t)&=& \frac{a^a}{\Gamma (a)}\, t^{a-1}\, e^{-at}\1_{[0 ,\infty )} (t)\nonumber\\
f_\delta^P (t)&=&\frac{1}{\delta (1-\delta )}\, (\delta /t)^{(2-\delta)/(1-\delta )}\1_{[\delta ,\infty )} (t)\ ,
\label{eq:fgamma}
\end{eqnarray}
both parametrized in order to have normalized mean with $a>0$ and $\delta\in (0,1)$. 
At jump times $\tau_x^k$, if there is a particle at site $x$ it jumps to $x+1$ provided this site is currently empty, otherwise the configuration $\eta$ is not updated. In formulae,
\begin{equation}
\mbox{if }t=\tau_x^k \mbox{ and }\eta_x (t-)\big( 1-\eta_{x+1} (t-)\big) =1\quad\mbox{then}\quad \eta (t)=\eta (t-)^{x,x+1}
\label{eq:dynamics}
\end{equation}
for all $x\in\Lambda$ and $k\geq 1$. Here we use the common notation $\eta (t-)=\lim_{s\nearrow t} \eta (s)$ for the configuration just before the jump, and $\eta^{x,x+1} =\eta -\delta_x +\delta_{x+1}$ for the configuration where a particle moved from $x$ to $x+1$. The model can be defined analogously with jump processes $\tau_i$ attached to particles with index $i$ rather than sites. Note that, unless we are in the Markovian case, this is not equivalent to the site-based model. The latter exhibits a particle-hole symmetry and is simpler to analyze, so will be our main focus in the following with the particle-based model discussed briefly at the end.

A standard way to quantify the dispersion of samples of a probability distribution is the coefficient of variation $\chi$, defined as the ratio of standard deviation and expectation. Since we normalize the expectation to $1$, we have
\begin{equation}
\chi =\frac{\sqrt{\langle T^2\rangle -1}}{1} \quad\mbox{and}\quad\chi_\delta =1-\delta
\label{eq:chi}
\end{equation}
for the family (\ref{eq:fdelta}) of delayed exponentials. This equals $1$ for the Markovian case and vanishes as $\delta\to 1$ corresponding to a deterministic update after one unit of time. For heavy-tailed Pareto distributions this can also be larger than $1$ as is mentioned later in the discussion. Another very important characteristic of renewal processes is the residual lifetime
\begin{equation}
T_r =\lim_{t\to\infty} \min_{k\geq 1}\{ \tau^k -t :\tau^k >t\}\ ,
\label{eq:restime}
\end{equation}
which is the long-time limit of the remaining time until the next jump attempt. Intuitively, this corresponds to the time until the next jump when observing the process at a `random' time, which is equivalent to a stationary observation of this quantity. The distribution of $T_r$ can be computed explicitly \cite{KarlinTaylor} using (\ref{eq:dwelldist}), 
\begin{equation}
\P (T_r \leq t)=\frac{1}{\langle T\rangle} \int_0^t (1-F(s))\, ds\ .
\label{eq:resdist}
\end{equation}
For exponential $T$ this is simply equal to the waiting time distribution $F(t)=1-e^{-t}$, corresponding to the memory-less property of Markov processes. For the distribution (\ref{eq:fdelta}) we get for the expected residual lifetime
\begin{equation}
\langle T_r \rangle_\delta =1-\delta +\frac12 \delta^2 \ ,
\label{eq:res}
\end{equation}
with $\langle T_r \rangle_{\delta =0}=1$ in the Markovian case as expected. In the deterministic limit $\delta\to 1$ this converges to the minimal value $1/2$, corresponding to the average of a uniform random variable on $[0,1]$ which is the interval between two jump events. For any mean-$1$ distribution with positive variance the expected residual lifetime is larger than $1/2$. It can also be larger than the Markovian value $1$ or even diverge due to heavy tails, as is discussed later for Pareto distributions. Note that, since we standardize the mean,
\begin{equation}
\mathrm{Var} (T)=2\langle T_r \rangle -1
\label{eq:varres}
\end{equation}
and the variation coefficient $\chi$ and $\langle T_r \rangle$ are not independent.  However, it is still useful to consider both in order to predict the shape of fundamental diagrams.

\begin{figure}%
\begin{center}
\mbox{\includegraphics[width=0.5\textwidth]{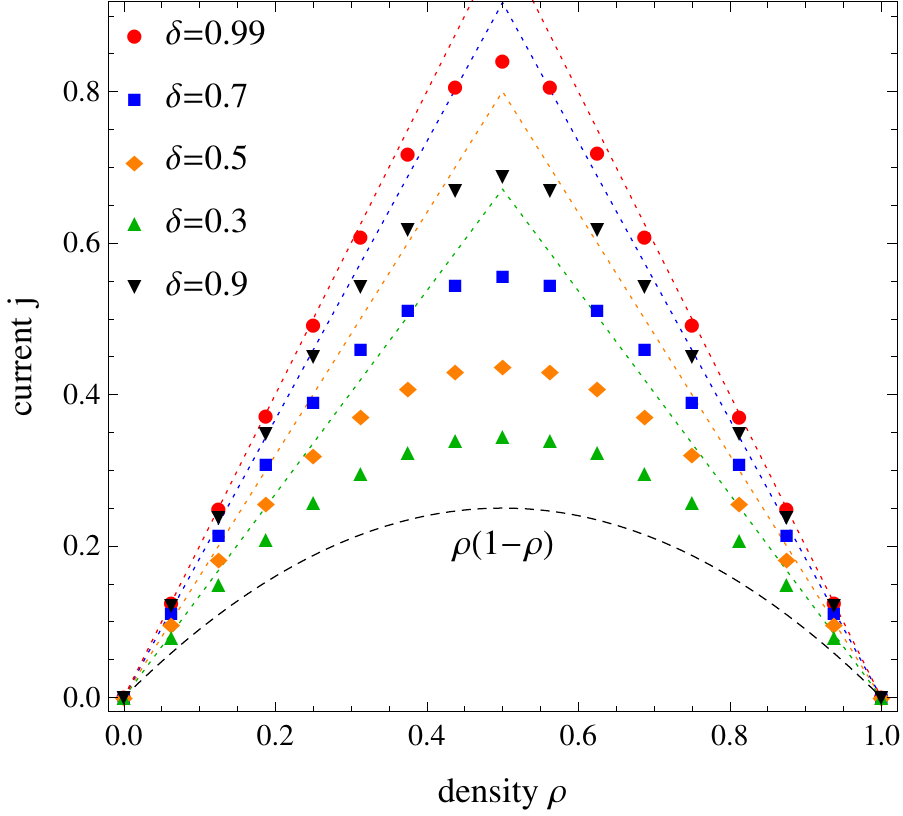}\quad
\includegraphics[width=0.45\textwidth]{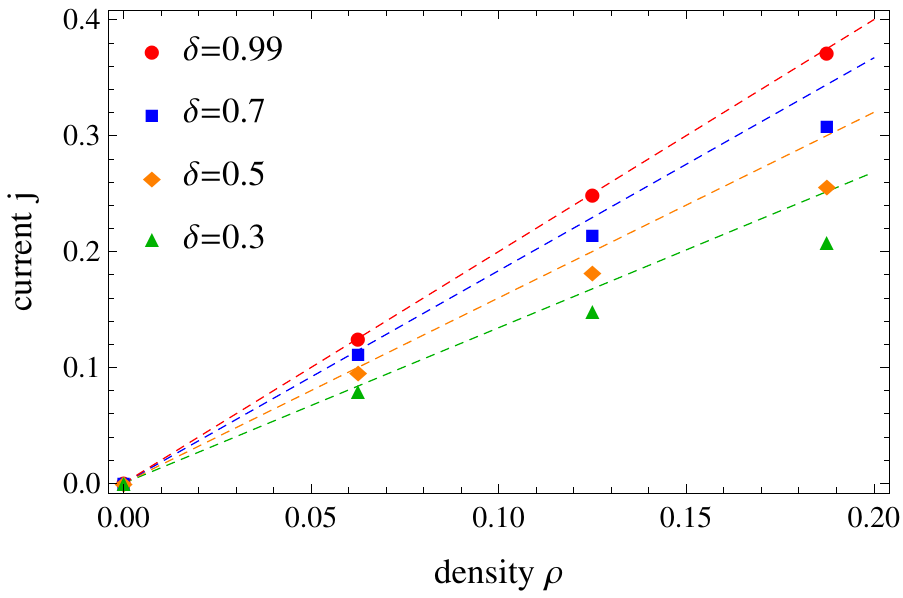}}
\end{center}
\caption{Fundamental diagrams (\ref{eq:jrho}) for the family of waiting time distributions (\ref{eq:fdelta}). The current is clearly enhanced compared to the Markovian case (\ref{eq:fundtasep}), and limiting slopes at high and low densities are given by inverse expected residual lifetimes (\ref{eq:res}), see also the zoomed in version on the right. In the limit of deterministic waiting times $\delta\to 1$, the current approaches the maximal triangular shape (\ref{eq:result1}). Data are from MC simulations with system size $L=1024$ and error bars are of the size of the symbols.}%
\label{fig:fd}%
\end{figure}

In general, to avoid degeneracies we assume the renewal processes $\tau_x$ to be stationary and ergodic, i.e., waiting times have a continuous distribution with positive variance, and the time of the first jump attempt $\tau_x^1$ is distributed as $T_r$ independently for each $x$, to ensure stationarity. 
We denote by
\begin{equation}
J_L (t) :=\frac{1}{L}\sum_{x\in\Lambda}\sum_{k:\tau_x^k \leq t} \eta_x (\tau_x^k )\big( 1-\eta_{x+1} (\tau_x^k )\big)
\label{eq:jlt}
\end{equation}
the number of successful jumps per lattice site up to time $t>0$. Then, if we fix $N$ particles on the periodic lattice with $L\geq N$ sites, we have
\begin{equation}
j(L,N):=\lim_{t\to\infty} \frac{1}{t} J_L (t)=\E^\delta_{L,N} [J_L (1)]
\label{eq:jln}
\end{equation}
for the average stationary current. Here the expectation is w.r.t.\ the time evolution of the process started in the stationary state, and convergence holds as a result of the standard renewal theorem (see e.g.,~\cite{KarlinTaylor}). We are interested in the thermodynamic limit
\begin{equation}
j(\rho )=\lim_{L,N\to\infty\atop N/L\to\rho} j(L,N)\ , 
\label{eq:jrho}
\end{equation}
and in the Markovian case this is simply given by the stationary expectation of the jump rate. For translation invariant systems, the stationary state $P^{\delta =0}_\rho$ is known to factorize with Bernoulli marginals $\rho =P^0_\rho (\eta_x =1 )=1-P^0_\rho (\eta_x =0 )$ (see e.g.,~\cite{Derrida1998}). Therefore
\begin{equation}
j(\rho )=\langle \eta_x (1-\eta_{x+1}) \rangle_\rho =\rho (1-\rho )\quad\mbox{for}\quad\delta =0\ ,
\label{eq:fundtasep}
\end{equation}
which is the established fundamental diagram for the standard TASEP. In the non-Markovian case, the stationary state $P_\rho^\delta$ for $\delta >0$ exhibits non-trivial correlations (cf.\ figure~\ref{fig:tagged}), and analytic characterizations of the $P_\rho^\delta$ are not available to our knowledge. Existence and uniqueness of stationary states, and in particular ergodic convergence, are interesting mathematical problems and current work in progress. Heuristically, a temporal correlation length $\theta$ induced by the waiting time distribution is proportional to the inverse coefficient of variation. With normalized mean we get
\begin{equation}
\theta =\frac{1}{\chi} \quad\mbox{and}\quad\theta_\delta=\frac{1}{1-\delta} ,
\label{eq:theta}
\end{equation}
for the distribution (\ref{eq:fdelta}), which is equal to $1$ in the Markovian case and diverges as $\delta\to 1$. Since these time correlations are finite, we expect the process to be ergodic and to converge to the unique stationary distribution $P_\rho^\delta$. Indeed, measurements of fundamental diagrams from Monte Carlo simulations shown in figure \ref{fig:fd} support this.

\section{Main results\label{s:results}}

In the limit of low and high density, the slope of the fundamental diagram can be understood by the expected jump rate of a single particle or vacancy, respectively. In large systems $L\gg\theta$ the arrival time of an isolated particle on a site $x$ is effectively independent of the renewal process $\tau_x$. So the holding time for the particle at this site will be given by the residual lifetime $T_r$, leading to an average jump rate of $1/\langle T_r\rangle$. This is confirmed in figure \ref{fig:fd} for the distributions (\ref{eq:fdelta}), where we see that
\begin{eqnarray}
j'_\delta (\rho )&\to &\pm 1/\langle T_r\rangle_\delta\quad\mbox{as }\rho\to 0\mbox{ or }1,\mbox{ respectively},\quad\mbox{and}\nonumber\\
j_\delta (\rho )&\to & 2\min\{ \rho ,1-\rho\}\quad\mbox{as }\delta\to 1\ .
\label{eq:result1}
\end{eqnarray}
In the deterministic limit $\delta\to 1$ the fundamental diagram approaches the limiting triangle. In this case, there are no blocked jump attempts for densities $\rho\leq 1/2$ since particles are separated by holes and move in synchrony like a single particle (cf.\ figure~\ref{fig:tagged}). For densities $\rho\geq 1/2$ the same argument holds for holes by symmetry of the dynamics and particle positions are highly correlated so as to maximize the current. All simulation results have been checked to be robust in the system size and we keep $L=|\Lambda |=1024$ for all simulation data presented.

\begin{figure}%
\begin{center}
\mbox{\includegraphics[width=0.47\textwidth]{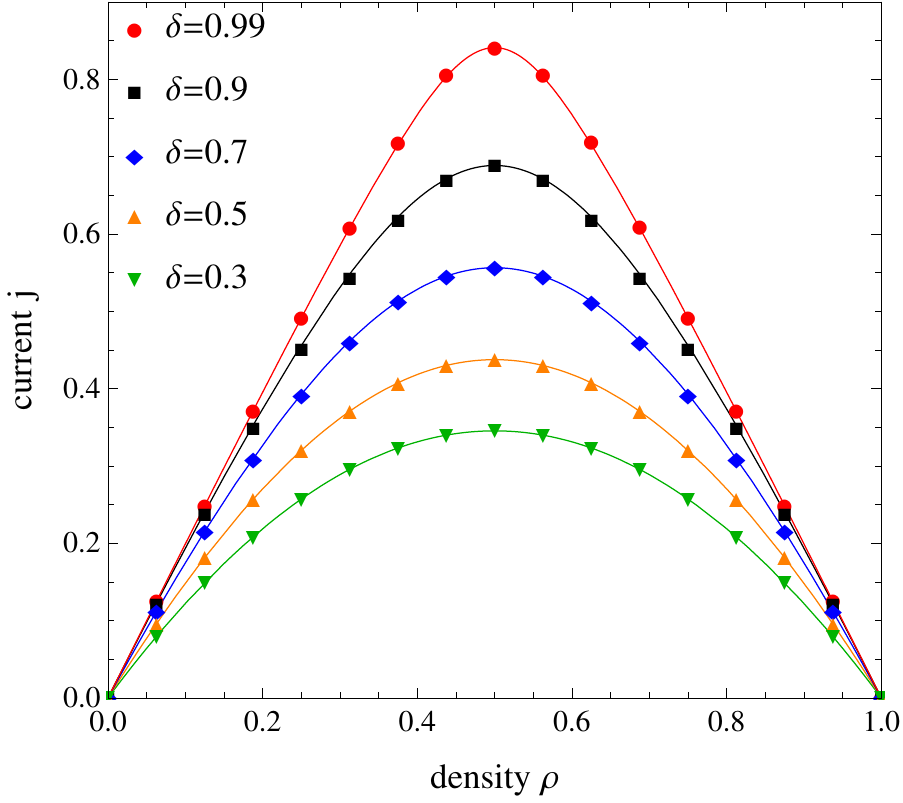}\quad
\includegraphics[width=0.47\textwidth]{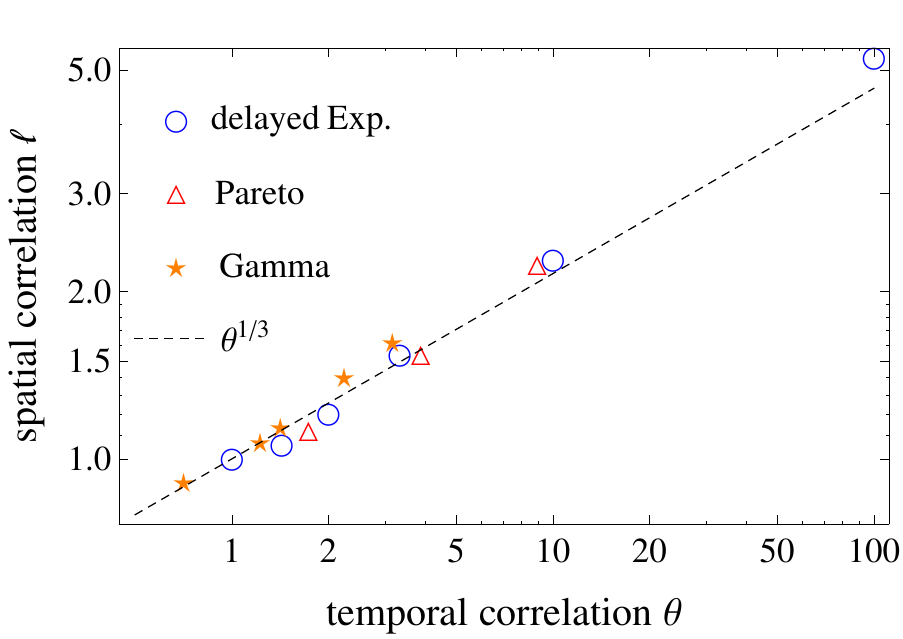}}
\end{center}
\caption{Fundamental diagrams for different waiting time distributions show excellent agreement with the theory (\ref{eq:gcfund}) after fitting the maximal current, as shown on the left for the family of delayed exponentials (\ref{eq:fdelta}). The maximal current can be predicted without fitting from the KPZ scaling relation (\ref{eq:scaling}) of the temporal and spatial correlation lengths $\theta$ (\ref{eq:theta}) and $\ell$ (\ref{eq:scorr}). Fitted values of $\ell$ follow this relation very well as shown on the right. Data are from MC simulations with system size $L=1024$ and error bars are of the size of the symbols.}%
\label{fig:fdpa}%
\end{figure}

\begin{figure}%
\begin{center}
\mbox{\includegraphics[width=0.47\textwidth]{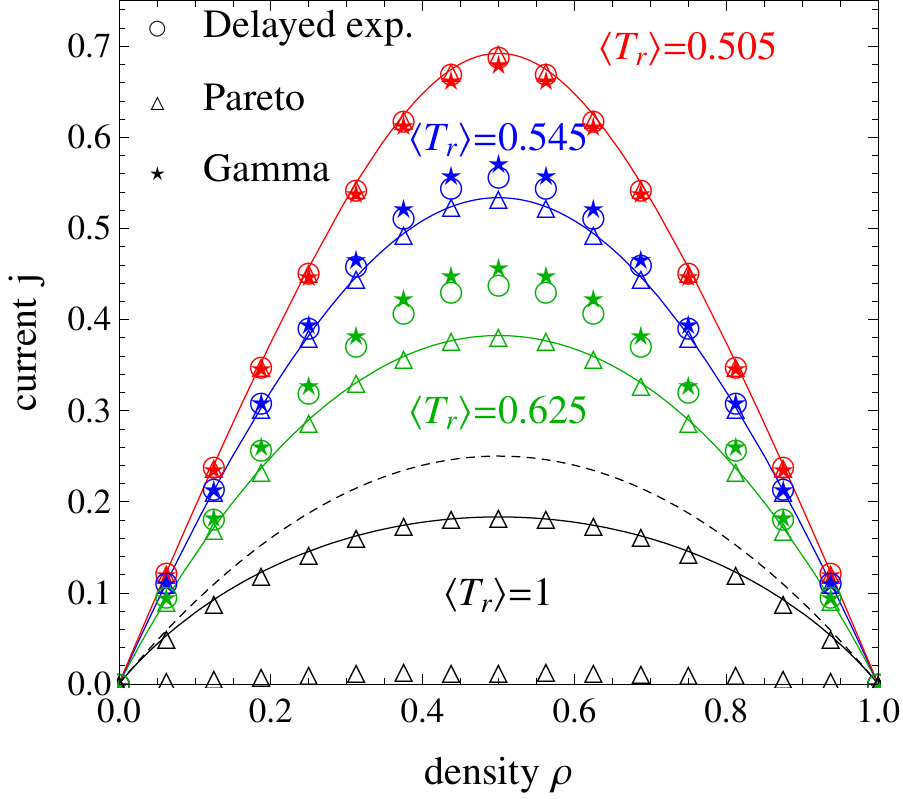}\quad
\includegraphics[width=0.47\textwidth]{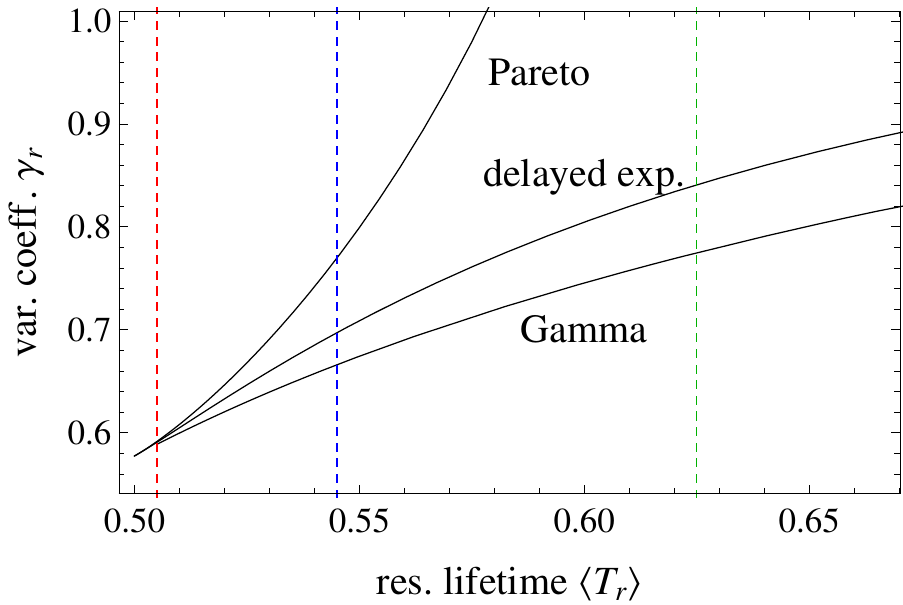}}
\end{center}
\caption{Comparison of fundamental diagrams for delayed exponential, Pareto and Gamma distributions with the same expected residual lifetimes $\langle T_r \rangle$. Theoretical predictions are fitted to Pareto waiting times, which lead to lower currents. This is also illustrated comparing to the Markovian case (dashed black line) for $\langle T_r \rangle =1$, and the approach (\ref{eq:gcfund}) still gives excellent agreement for fits with $\beta <0$ (full black line).  For Pareto with $\delta \leq 0.5$ currents essentially vanish, since $\langle T_r \rangle$ diverges (\ref{eq:partr}), data shown are for $\delta =0.3$. Right: Deviations between different distributions can be explained qualitatively by higher order statistics such as the variation coefficient $\gamma_r$ of the residual lifetime, as explained in (\ref{eq:vartr}). Larger $\gamma_r$ indicates a reduced maximal current due to higher fluctuations. Dashed vertical lines mark the values of $\langle T_r \rangle$ used on the left providing good qualitative agreement. Data are from MC simulations with system size $L=1024$ and error bars are of the size of the symbols.}%
\label{fig:fdvar}%
\end{figure}

For intermediate range values $\delta\in (0,1)$ we expect a competition between current maximization and entropic fluctuations reminiscent of the classical paradigm of Gibbs measures in statistical mechanics, with the current playing the role of negative energy. In the site-based process, maximization of the current $J_L (1)$ over a unit time interval (\ref{eq:jlt}) is essentially equivalent to maximizing the expected instantaneous current given by
\begin{equation}\label{eq:opti}
\Jcal (\eta )=\sum_{x\in\Lambda} \eta_x (1-\eta_{x+1} )\ ,
\end{equation}
which is simply the total jump rate in the Markovian case. Therefore, we can estimate the fundamental diagram by a stationary expectation w.r.t.\ $P_\rho^\delta$ without taking holding times into account. This is not possible for the particle-based model as we discuss later. Note that we omit the prefactor $1/\langle T_r\rangle_\delta$ in the function $\Jcal$ and focus only on the $\eta$-dependent part, since this simplifies the presentation. Our main hypothesis is that we can approximate a grand-canonical version of $P_\rho^\delta$ by a tilt $P^\beta_\mu$ of the grand-canonical Markovian distribution $P_\mu^0$, where the chemical potential $\mu\in\R$ is the usual conjugate parameter to $\rho$. Then we have
\begin{equation}
P^\beta_\mu (\eta )=\frac{1}{\langle e^{\beta \Jcal}\rangle_\mu^0} e^{\beta \Jcal (\eta )}P_\mu^0 (\eta )\ ,
\label{eq:tilt}
\end{equation}
where $\beta\geq 0$ is the effective inverse `temperature' conjugate to $\delta$, and $\langle ..\rangle_\mu^0$ denotes expectation w.r.t.~the distribution $P_\mu^0$. In other words, we use the known $P_\mu^0$ as the reference distribution at infinite temperature ($\beta =0$), and for finite temperature we re-weight configurations according to their instantaneous current (\ref{eq:opti}). 
The associated free energy
\begin{equation}
f(\beta ,\mu )=\lim_{L\to\infty}\frac{1}{L}\log\langle e^{\beta \Jcal }\rangle_\mu^0
\label{eq:free}
\end{equation}
can be computed using the transfer matrix $M={1\ \ 1\choose e^{\beta +\mu}\ e^\mu}$, see e.g.\ \cite{peliti}, Section 5.1.  Specifically, $\langle e^{\beta \Jcal }\rangle_\mu^0$ is given by the trace $\mathrm{Tr}(M^L )$ and hence, taking the $L \to \infty$ limit, the free energy is determined by the largest eigenvalue of $M$ as
\begin{equation}
f(\beta ,\mu )=\log \frac {(1+e^\mu )+\sqrt{(1+e^\mu )^2 +4(e^{\beta +\mu} -e^\mu )}}{2} \ .
\label{eq:free2}
\end{equation}
The stationary current and the density as a function of the chemical potential are then given by
\begin{equation}
j_\beta (\mu ) =\frac{1}{\langle T_r \rangle_\delta}\partial_\beta f(\beta ,\mu)\quad\mbox{and}\quad \rho_\beta (\mu )=\partial_\mu f(\beta ,\mu)\ .
\label{eq:gcfund}
\end{equation}
Evaluation provides lengthy formulae that can also be solved for $j_\beta (\rho )$ which we do not display here. The maximal current at density $1/2$ (corresponding to $\mu =0$) is given by the simple expression
\begin{equation}
j_\beta (0)=\frac{1}{\langle T_r \rangle_\delta 2(1+e^{-\beta /2})}\ ,
\label{eq:mc}
\end{equation}
and it remains to fix the parameter $\beta$ as a function of the delay $\delta$. Fitting $\beta$ with (\ref{eq:mc}) to simulation data provides a perfect match for the full fundamental diagram, as is shown in figure \ref{fig:fdpa} (left) for delayed exponential and in figure \ref{fig:fdvar} (left) for Pareto distributions. 

The observed fit values can be supported by the following heuristics. The correlation dependent part of the maximal current is given by the correlation function
\begin{equation}
p=\big\langle\eta_x (1-\eta_{x+1})\big\rangle_0^\beta =\langle T_r \rangle_\delta j_\beta (0)\ ,
\label{eq:}
\end{equation}
taking values in $[1/4,1/2)$. Properly normalized and interpreted as the parameter of a geometric distribution counting the number of correlated neighbouring sites, this induces a spatial correlation length
\begin{equation}
\ell =\frac{2p}{(1-2p)}=\frac{(1+e^{-\beta /2})^{-1}}{1-(1+e^{-\beta /2})^{-1}}=e^{\beta /2}
\in [1,\infty )\ .
\label{eq:scorr}
\end{equation}
Since the spatial and the temporal correlation length $\theta$ (\ref{eq:theta}) are finite, the large scale behaviour of the non-Markovian generalization should exhibit KPZ scaling like the usual TASEP \cite{kpz86,krug91}. The correlation lengths should therefore be related as
\begin{equation}
\ell \sim\theta^{1/3}\ ,
\label{eq:scaling}
\end{equation}
which is confirmed to good approximation in figure \ref{fig:fdpa} (right) for different distributions. Since the temporal correlation $\theta$ is fully determined by properties of the renewal process, this relation (with normalization given by the Markovian case) provides a method to predict the parameter $\beta$ which characterizes the spatial correlation, without having to fit to data. Note that an increase in spatial correlation lengths in general also leads to increased finite size-effects. However, since we consider periodic boundary conditions and the current is a self-averaging observable, those effects are still negligible for the system size of $L=1024$ which we present simulation data for.

Due to (\ref{eq:varres}) the residual lifetime $\langle T_r \rangle$ fully determines the temporal correlation length (\ref{eq:theta}) for waiting time distributions with standardized mean. So if (\ref{eq:scaling}) were to hold strictly, the fundamental diagrams should depend only on the first and second moment of the waiting time distribution and therefore be fully determined by $\langle T_r \rangle$. While this is true close to the deterministic limit where higher order cumulants of the waiting time distribution become negligible, we can see in figure \ref{fig:fdvar} (left) that for larger variation coefficients this is not the case (as to be expected). The larger the higher order moments, the lower we expect the maximal current to be due to enhanced fluctuations. This is confirmed in figure \ref{fig:fdvar} (right) where we plot the variation coefficient,
\begin{equation}
\gamma_r =\sqrt{\mathrm{Var}(T_r )}/\langle T_r \rangle,
\label{eq:vartr}
\end{equation}
of the residual lifetime as a function of the mean $\langle T_r \rangle$ for different distributions. $\gamma_r$ corresponds to a particular moment of third order, and we see that it is inversely correlated to the maximal current, which takes the lowest values for Pareto distributions. Close to the deterministic limit with $\langle T_r \rangle =0.505$ the fundamental diagrams and $\gamma_r$ essentially coincide for all the distributions.

\section{Discussion\label{s:discussion}}

\begin{figure}%
\begin{center}
\mbox{\includegraphics[width=0.47\textwidth]{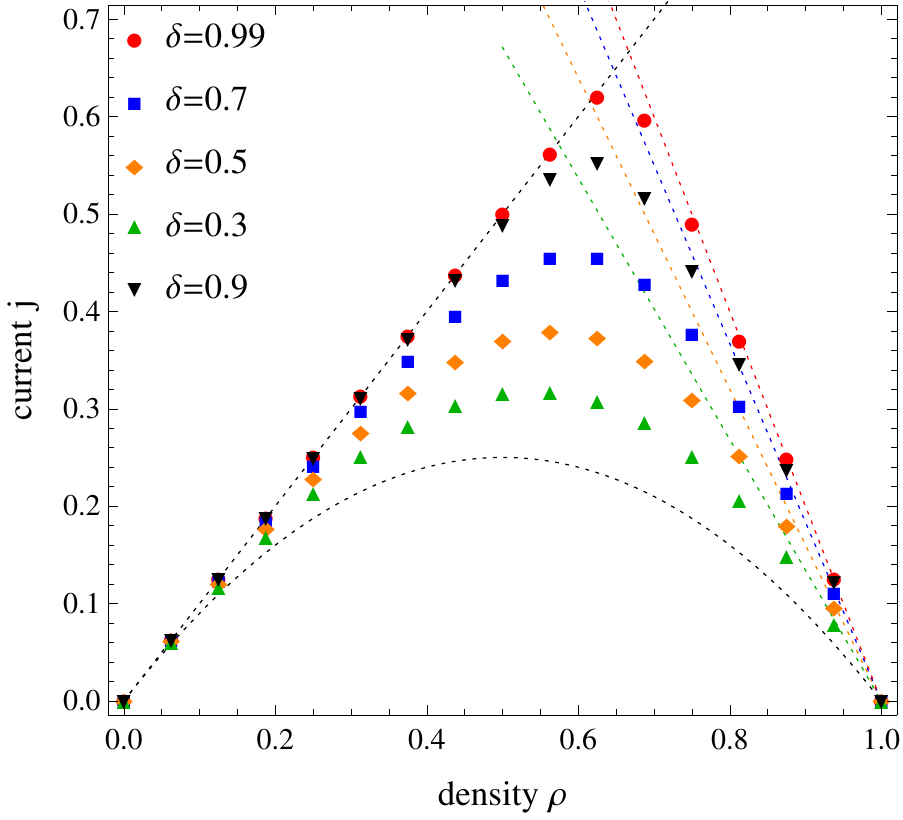}\quad
\includegraphics[width=0.47\textwidth]{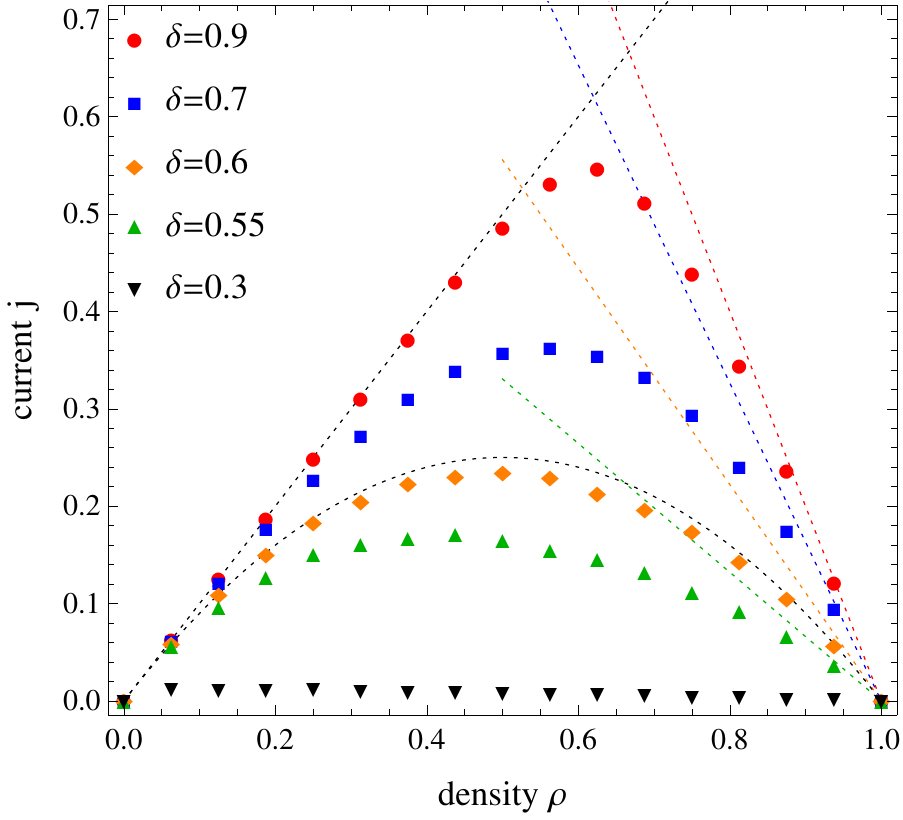}}
\end{center}
\caption{
Fundamental diagrams for particle-based dynamics for the family of delayed exponential distributions (\ref{eq:fdelta}) (left panel) and Pareto distributions (\ref{eq:fgamma}) (right panel). The asymptotic behaviour (\ref{eq:result1b}) indicated by dashed lines is confirmed well by simulation data, and the diagrams are no longer symmetric around $\rho =1/2$ due to loss of particle-hole symmetry. For Pareto distributions with low values of $\delta$ the current can decrease compared to the Markovian case (black dashed curve), and essentially vanishes for $\delta \leq 0.5$ up to finite size effects. Data are from MC simulations with system size $L=1024$ and error bars are of the size of the symbols.
}%
\label{fig:corr}%
\end{figure}

Even though our main focus of presentation was on enhanced current for waiting time distributions which are more regular than the exponential distribution, our theory also works well for reduced currents due to heavy tails. Using the general formula (\ref{eq:resdist}) we can compute
\begin{equation}
\langle T_r \rangle_\delta = \delta^2/(4\delta-2)\quad\mbox{for Pareto distributed $T$ given in }\ (\ref{eq:fgamma})\ ,
\label{eq:partr}
\end{equation}
analogously to (\ref{eq:res}). Note that for $\delta < 2-\sqrt{2}$, $\langle T_r \rangle$ is greater than the Markovian value of 1, leading to a suppression of the current (negative values of $\beta$). In fact, due to the heavy tail, (\ref{eq:partr}) holds only for $\delta >1/2$ and $\langle T_r \rangle_\delta$ diverges for $\delta\leq 1/2$. According to our theory the fundamental diagram should essentially vanish for such parameter values, which is confirmed by the lowest data points on the left panel of figure \ref{fig:fdvar}. The current is greatly reduced due to jamming events which have been studied in detail in \cite{Concannon2011}. 
Already for $\delta$ close to but greater than $1/2$ the empirical average is increasingly dominated by extreme values of the sample resulting from large fluctuations of the current, which requires a much more careful averaging and increased sample size than for higher values of $\delta$. This already becomes visible, near the maximal current, for data with $\delta =0.55$  which we show in figure \ref{fig:corr} (right panel), where the error bars are still of the order of the symbol size. 

If the renewal processes of jump times $\tau_i$ are attached to particles with indices $i$, the system loses the particle-hole symmetry and fundamental diagrams are no longer symmetric. An isolated single particle will then jump with an average rate $1$, whereas a single hole will have average waiting time $\langle T_r\rangle$ completely analogously to the bond-based model. Our first main result (\ref{eq:result1}) on the asymptotic behaviour of the fundamental diagram can be directly adapted to give
\begin{eqnarray}
j'_\delta (\rho )&\to &\left\{\begin{array}{cl} 1&,\ \mbox{as }\rho\to 0\\ -1/\langle T_r\rangle_\delta &,\ \mbox{as }\rho\to 1\end{array}\right.\ ,\quad\mbox{and}\nonumber\\
j_\delta (\rho )&\to &\min\{ \rho ,2-2\rho\}\quad\mbox{as }\delta\to 1\ .
\label{eq:result1b}
\end{eqnarray}
This is well confirmed by simulation data as shown in figure \ref{fig:corr}. The maximal current at density $\rho =2/3$ for $\delta\to 1$ can alternatively be explained by realizing that the update order of the particles will be frozen on a very long time scale. Therefore particles can form long-lived blocks where they move essentially in parallel one after the other within one unit of time. To achieve maximal current, these blocks have to be separated by an empty site. Since the update order is a uniform permutation due to independence of the renewal processes, the average block length is $2$ which leads to a maximal current at density $2/3$. In this limit our model is equivalent to the TASEP with frozen shuffle update studied in \cite{Appert-Rolland2011}. Note that in the bond-based model the update order of neighbouring particles changes all the time even in the limit $\delta\to 1$, so the formation of long lived blocks is not possible. Due to these blocks in the particle-based model, our simplified ansatz for the stationary distribution $P_\rho^\delta$ cannot be applied and a more detailed analysis including jump times is necessary. One possible approach is to describe the dynamics of inter-particle distances as a zero-range process, which is a standard mapping for the Markovian TASEP (see e.g.\ \cite{kipnis}) that also applies for the particle-based non-Markovian model. The zero-range process is basically a system of queues in tandem, which have been studied for non-exponential arrival and service time distributions. This leads to approximate formulae in terms of Laplace transforms of jump time statistics \cite{bertsimas}; 
a thorough investigation of this approach is the subject of current work in progress. Note that, for the bond-based model, departure and arrival processes of consecutive queues would be correlated and results in \cite{bertsimas} do not apply.

In summary, we have demonstrated a method to derive macroscopic transport properties, in the form of the fundamental diagram, for non-Markovian exclusion processes. The asymptotic results (\ref{eq:result1}) involving residual lifetimes are of a very general nature and can be applied directly to other driven diffusive systems, including also partially asymmetric dynamics. The applicability of our ansatz (\ref{eq:tilt}) for the intermediate regime will depend on details of the model and may have to be adapted with an enlarged state space. However, the underlying principle of a description of correlated stationary states on the basis of maximization of the instantaneous current should apply in a very general sense. We have thus reported here a first step towards a broader understanding of the large-scale dynamics of non-Markovian driven diffusive systems which is highly relevant in applications and poses very interesting theoretical questions for future research.

\section*{Acknowledgments}
We thank Richard Blythe, Neil Jenkins and Gunter Sch\"utz for helpful discussions. This work was supported by the Engineering and Physical Sciences Research Council (EPSRC), Grant No. EP/E501311/1.

\section*{References}


\end{document}